# Electronic properties of the steps in bilayer *Td*-$WTe_2$

Mari Ohfuchi[1]*, Akihiko Sekine[1], Manabu Ohtomo[1], Kenichi Kawaguchi[1]

[1]*Fujitsu Research, Fujitsu Limited, Atsugi, Kanagawa 243-0197, Japan*

E-mail: mari.ohfuti@fujitsu.com

Monolayer $WTe_2$ stripes are quantum spin Hall (QSH) insulators. Density functional theory was used for investigating the electronic properties of the stripes and steps in bilayer *Td*-$WTe_2$. For the stripes oriented along the dimer chains of W atoms ($x$ direction), the hybridization between the two layers suppresses the QSH states. However, the QSH nature can be recovered by forming a step, depending on the atomic structure of the step. Conversely, the stripes and steps along the $y$ direction maintain the QSH states. These findings can expand the application range of the QSH states in $WTe_2$.

Since the discovery of isolated graphene,[1,2] the quest for identifying exceptional two-dimensional (2D) materials has continued, such as transition metal dichalcogenides (TMDCs). $WTe_2$ is a TMDC and its complex and interesting properties, similar to other 2D materials,[1–4] highly depend on the number of layers present in it. Bulk *Td*-$WTe_2$ is a semimetal with a space group 31 ($Pmn2_1$)[5–7] and is predicted as a type-II Weyl semimetal.[8] The balanced electron–hole population affords a large magnetoresistance.[7] Furthermore, when the Weyl points are annihilated via a strain or lattice distortion, bulk $WTe_2$ can become a higher-order topological insulator with topologically protected hinge states.[9–11] By contrast, several studies have reported that monolayer and bilayer $WTe_2$ are insulators with a small bandgap.[12–16] Recently, the observed bandgap is proposed to be caused by a Coulomb gap,[17] strain,[18] or exciton condensation[19] instead of the original energy band structure.[12,14,16]

Monolayer $WTe_2$ was predicted to be a quantum spin Hall (QSH) insulator during an early stage of the research.[20] The QSH state can be applied to spintronics and topological quantum computation.[21–23] Localized edge states in the bulk bandgap, supporting the existence of the QSH states, have been observed in the monolayers[13–15,24] and the step edges.[25] Conversely, bilayer $WTe_2$ has provided a platform for studying the nonlinear Hall effect caused by the inversion symmetry breaking.[26–30] In this study, we focus on bilayer $WTe_2$ and investigate how the QSH states are maintained and how the QSH states are recovered by forming a step when suppressed. This study can expand the application range of the QSH states in $WTe_2$.

Most previous studies investigating the edge states of $WTe_2$ are based on low-energy tight-binding models fitted to the density functional theory (DFT) band structure.[12,20,31,32] Moreover, the shape of the localized edge state bands associated with monolayer $WTe_2$ depends on the atomic structure of the edge.[32] Here, we focus on the changes in the localized edge state bands caused by the layer stacking without changing the atomic structure of the edges. We perform DFT calculations directly on the stripe or step geometries of bilayer $WTe_2$ to investigate the interactions between the two layers. We adopted the OpenMX code,[33–36] which has been successfully used for studying 2D materials, including $WTe_2$,[37–43] to perform DFT calculations. The exchange–correlation potential was treated with a generalized gradient approximation.[44] The electron–ion interaction was described using norm-conserving pseudopotentials, including spin-orbit coupling (SOC).[45,46] The pseudoatomic orbitals denoted by W7.0-s3p2d2 and Te7.0-s3p3d2 were used as the basis set. The energy band structure obtained for bulk *Td*-$WTe_2$ crystal structure[5] using these methods is consistent with previous results[7,8] (see supplementary data, Figs. S1 and S2, for the atomic structure and energy band diagram of bulk *Td*-$WTe_2$). The unit cell of *Td*-$WTe_2$ has two layers, and the W atoms of each layer form dimer chains along the *x* direction. The two layers are stacked with a misalignment, yielding the bulk noncentrosymmetric nature and undegenerated SOC split bands.

Figure 1 depicts the geometric and electronic properties for monolayer *Td*-$WTe_2$ stripes. The stripes along the *x* and *y* directions are referred to as *x*- and y-stripes and are presented in Figs. 1(a) and 1(d), respectively. The widths of the *x*- and *y*-stripes comprise ten- and twenty-unit cells, respectively, affording approximately the same width (6.2 nm for the *x*-stripe and 7.0 nm for the *y*-stripe). We used the atomic structures as a cut out from bulk *Td*-$WTe_2$, where we did not intend to break the dimer chains of W atoms along the *x* direction. Although the monolayer cut out from bulk *Td*-$WTe_2$ strictly does not have an inversion symmetry, the approximately degenerated SOC split bands in Figs. 1(b) and 1(e) demonstrate the existence of a "pseudoinversion" symmetry. Two pairs of edge state bands connect the valence and conduction bands for both stripes, consistent with the fact that the atomic structure of the *x*- and *y*-stripes is topologically continuous. Examining the spin densities [see supplementary data, Fig. S3, for the spin densities at the points indicated by broken gray lines in Figs. 1(b) and 1(e)], we can confirm the presence of two edge-localized helical states associated by a pseudoinversion symmetry with almost linear bands connected to the valence bands. These bands deviate from linearity and the spin density spreads inside the stripes near the valence bands. Figures 1(c) and 1(f) depict the local density of states (LDOS) at the edge and near the center of the stripe as the bulk of monolayer. The finite LDOS values in the bulk bandgap near the Fermi level, combined with the aforementioned results, provide the distinct characteristics of a QSH insulator. Adopting a hybrid functional may afford a wider bandgap than our results, as introducing a hybrid functional reportedly widens the bandgap of monolayer and bilayer sheets.[12,14,16)]

First, we discuss the *x*-stripes of bilayer *Td*-$WTe_2$. Even when the stacking alignment and width of the stripe are determined, two geometric models can be considered: (A) the atomic structure where the top right atoms overhangs, as depicted in Fig. 2(a) and (B) the atomic structure where the top left atoms overhangs [see supplementary data, Fig. S4(a)]. The energy band diagram and spin densities associated with the A-*x*-stripe are depicted in Figs. 2(b) and 2(c). We can see eight bands starting from around 0.3 eV at Γ point, which are described as follows. The edge-localized states of monolayer $WTe_2$ in Fig. 1(b) are split due to the breaking of the pseudoinversion symmetry and because of hybridization between two layers, affording bonding and antibonding states. As the wavenumber increases, the edge-localized state bands considerably bend and intersect, changing the hybrid weights in the spin densities, and are unlikely to be in the QSH state. The B-*x*-stripe can be similarly understood [see supplementary data, Figs. S4(b) and S4(c), for the results obtained from the B-*x*-stripe].

Next, we consider the *x*-steps in bilayer *Td*-$WTe_2$. Figure 3(a) [3(d)] is the atomic structure in which the lower right (left) is extended twice from the A-*x*-stripe of bilayer. Note that the atomic structure inverted after extending the upper right (left) twice is different from that depicted in Fig. 3(a) [3(d)]. We can generate four step geometries from the B-*x*-stripe. (See supplementary data, Fig. S6, for the six atomic structures except the two shown in Fig. 3). For a step on the right side, as shown in Fig. 3(b), the lower four edge-localized state bands (indicated by the broken gray lines) do not bend significantly or intersect each other. We can confirm that the spin density is not hybridized on the step side [see supplementary data, Fig. S5(a)]. The LDOS at the edges and near the center of each layer are shown in Fig. 3(c). The LDOS spectra in the two central areas of each layer (areas 10 and 11 for the lower layer, and areas 25 and 26 for the upper layer) almost overlap and seemingly form a gap near the Fermi level. We can find the finite LDOS values at the edges in the gap of each layer. Based on these results, we conclude that the QSH states are recovered on the step side. Conversely, for a step on the left side, as shown in Fig. 3(e), the edge-localized state bands exhibit the same appearance as the *x*-stripe of the bilayer [Fig. 2(b)]. The spin densities are still hybridized between two layers on the step side [see supplementary data, Fig. S5(b)]. This phenomenon can be further confirmed in the LDOS spectrum shown in Fig. 3 (f). The LDOS in area 11 around the Fermi level is larger than that in area 10. Unlike the step on the right side, the QSH states remain suppressed. For the six remaining atomic structures, similar band diagrams were obtained depending on whether the step is on the right or left side (see supplementary data, Fig. S7, for the energy band diagrams). This difference seems to be due to the atomic structure at the edge of the upper layer of the step relative to the lower layer. In other words, to recover a QSH state, the upper edge of the step should not have an atomic structure whose edge state hybridizes to the state of the underlying layer.

Finally, we discuss the *y*-direction stripes and steps in bilayer *Td*-$WTe_2$. The atomic structure for the *y*-stripe is depicted in Fig. 4(a). The two opposite zigzags are closed at the right edge and open at the left edge. The corresponding energy band diagram is presented in Fig. 4(b). We can see four bands with almost linear dispersion connected to the valence bands (indicated by the broken gray lines). The spin densities in Fig. 4(c) show that each state on these bands spreads evenly over the two layers and is connected to the bulk state without changing the hybrid weight. These features indicate that these four bands are in the QSH state. The energy difference between the two states at the right edge is larger than that at the left edge. The interaction between the two layers is strong on the right side because the two opposite zigzags are closed, explaining the difference in obtained results when introducing steps to the right and left edges. Figure 5(a) [5(d)] is the atomic structure where the lower right (left) is extended twice. For a step on the right side, as shown in Fig. 5(b), the four bands are almost overlapping. The LDOS spectra on the step side presented in Fig. 5(c) form a single peak between 0 and 0.1 eV. Conversely, when there is a step on the left side, the four bands have the same energy spread as the *y*-stripe [Fig. 5(e)]. Nevertheless, only one peak between 0 and 0.1 eV is observed in the LDOS spectra on the step side [Fig. 5(f)]. These results show that in either case, the two QSH states on the step side do not interact and are as robust as the QSH states of monolayer $WTe_2$.

To summarize, we have investigated the electronic properties of the stripes and steps in bilayer *Td*-$WTe_2$ using DFT calculations. For the *x*-stripes, the hybridization between two layers suppresses the QSH states. However, the QSH nature can be recovered by forming a step, depending on the atomic structure of the step. Conversely, the *y*-stripes and *y*-steps maintain the QSH edge states. The QSH states on the step side are as robust as the QSH states of monolayer $WTe_2$. We believe that these findings will expand the application range of the QSH states of $WTe_2$ and help design the device applications and interpret the experiments on the electronic properties of the steps in bilayer $WTe_2$.

**Acknowledgments**

We would like to thank Dr. S. Sato, Dr. Y. Doi, Dr. J. Yamaguchi, and Dr. M. Hosoda for their advice, encouragement, and support. We thank the Research Institute for Information Technology, Kyusyu University, Japan for providing access to the supercomputer system for performing the calculations in this study.

## Figure Captions

**Fig. 1.** Geometric and electronic properties of monolayer *Td*-$WTe_2$ stripes: atomic structure, energy band diagram, and local density of states (LDOS) for (a) (b) (c) the *x*-stripe and (d) (e) (f) the *y*-stripe, respectively. The atomic structures in (a) and (d) are projected toward the *x* and *y* directions, respectively. The orange and green spheres denote W and Te atoms, respectively. The energy bands are drawn alternately in blue and red in ascending order of the energy. The Fermi level is set to zero. The areas for the LDOS in (c) and (f) are indicated in (a) and (d), respectively. [See supplementary data, Fig. S3, for the spin densities at the points indicated by the broken gray lines in (b) and (e)].

**Fig. 2.** (a) Atomic structure, (b) energy band diagram, and (c) spin densities for the A-*x*-stripe of bilayer $WTe_2$. The atomic structure is projected toward the *x* direction. The orange and green spheres denote W and Te atoms, respectively. The broken red line represents the alignment of layer stacking. The energy bands are drawn alternately in blue and red in ascending order of the energy. The spin densities are presented for the points indicated by the broken gray lines in (b). The numerical values show the wavenumber and energy at each point. The blue and red represent up and down spins, respectively.

**Fig. 3.** Results of the *x*-steps in bilayer *Td*-$WTe_2$: atomic structure, energy band diagram, and local density of states (LDOS) for the (a) (b) (c) right and (d) (e) (f) left steps, respectively. The atomic structures are projected toward the *x* direction. The orange and green spheres denote W and Te atoms, respectively. The red lines represent the alignment of layer stacking. The energy bands are drawn alternately in blue and red in ascending order of the energy. The Fermi level is set to zero. The areas for the LDOS in (c) and (f) are indicated in (a) and (d), respectively. [See supplementary data, Fig. S6, for the spin densities at the points indicated by the broken gray lines in (b) and (e)].

**Fig. 4.** (a) Atomic structure, (b) energy band diagram, and (c) spin densities for the *y*-stripe of bilayer $WTe_2$. The atomic structure is projected toward the *y* direction. The orange and green spheres denote W and Te atoms, respectively. The energy bands are drawn alternately in blue and red in ascending order of the energy. The Fermi level is set to zero. The spin densities at the points indicated by broken gray lines in (b). The numerical values show the wavenumber and energy at each point. The blue and red represent up and down spins, respectively.

**Fig. 5.** (a) (d) Atomic structures of the *y*-steps in bilayer *Td*-$WTe_2$. The orange spheres denote W atoms, and the green spheres denote Te atoms, projected toward the *y* direction. (b) (e) Energy band diagrams

for (a) (d). The energy bands are drawn alternately in blue and red in ascending order of the energy. The Fermi level is set to zero. (c) (f) Local density of states (LDOS) for (a) (d). The areas for LDOS in (c) and (f) are indicated in (a) and (d), respectively.

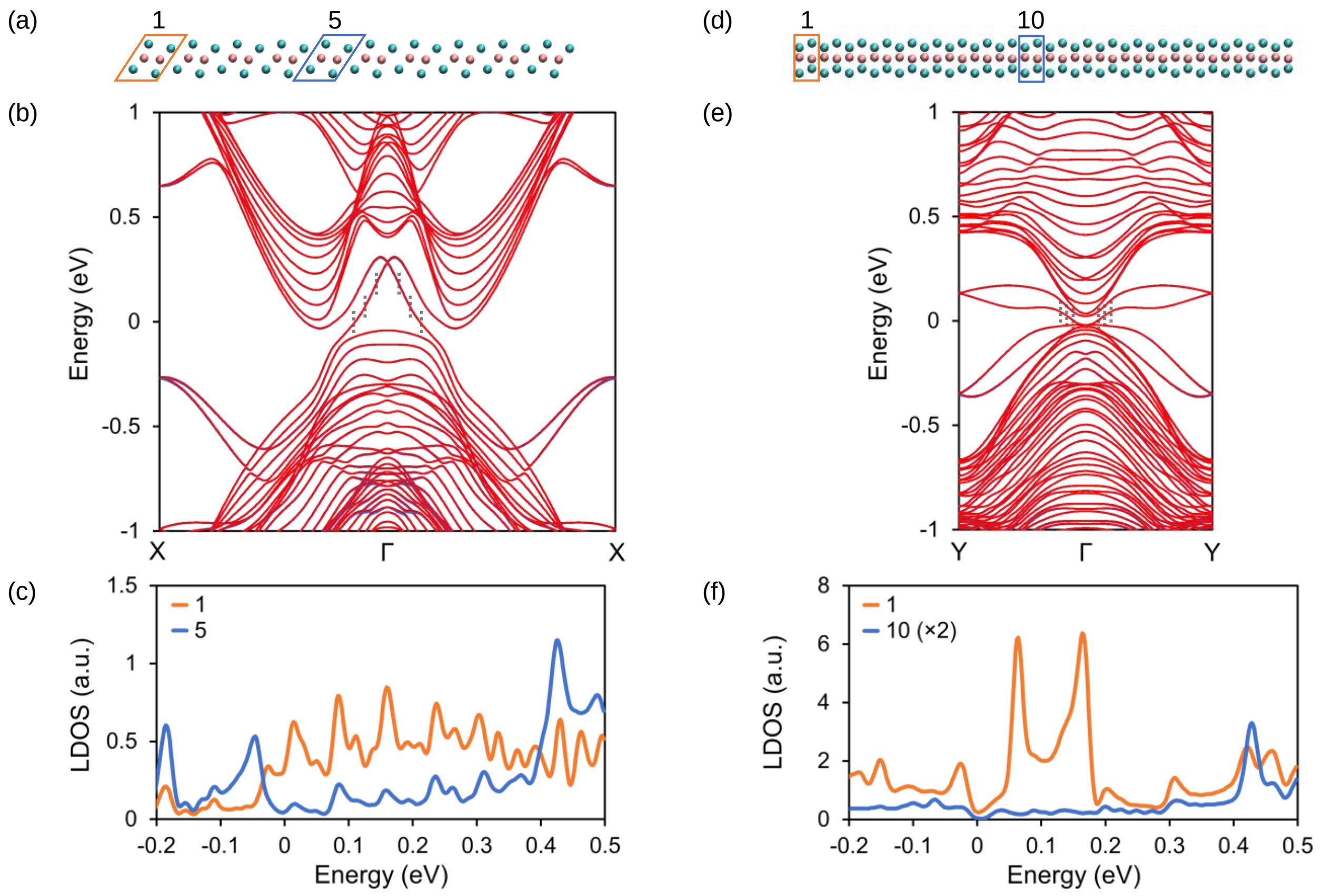

Fig.1.

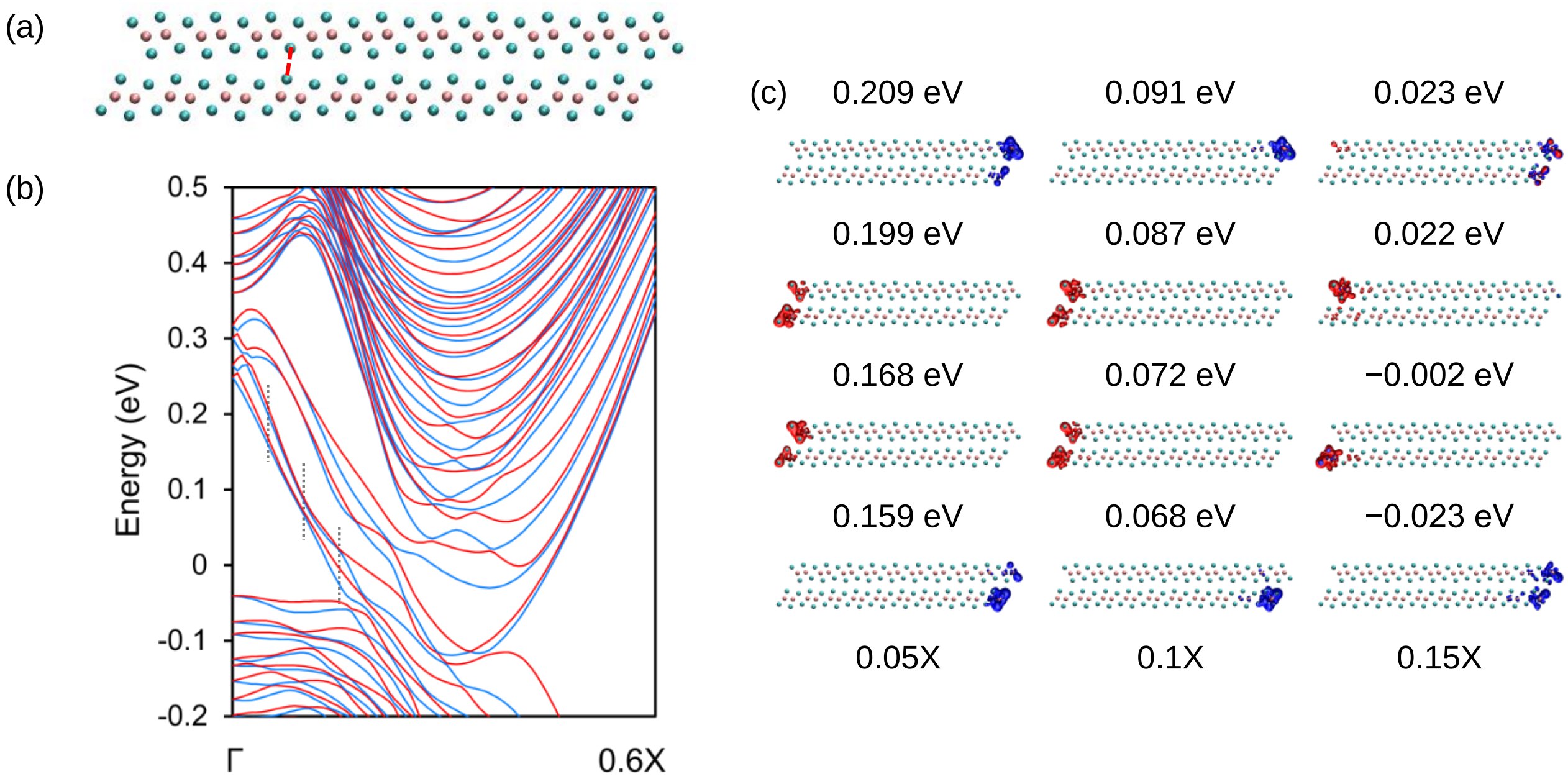

Fig.2.

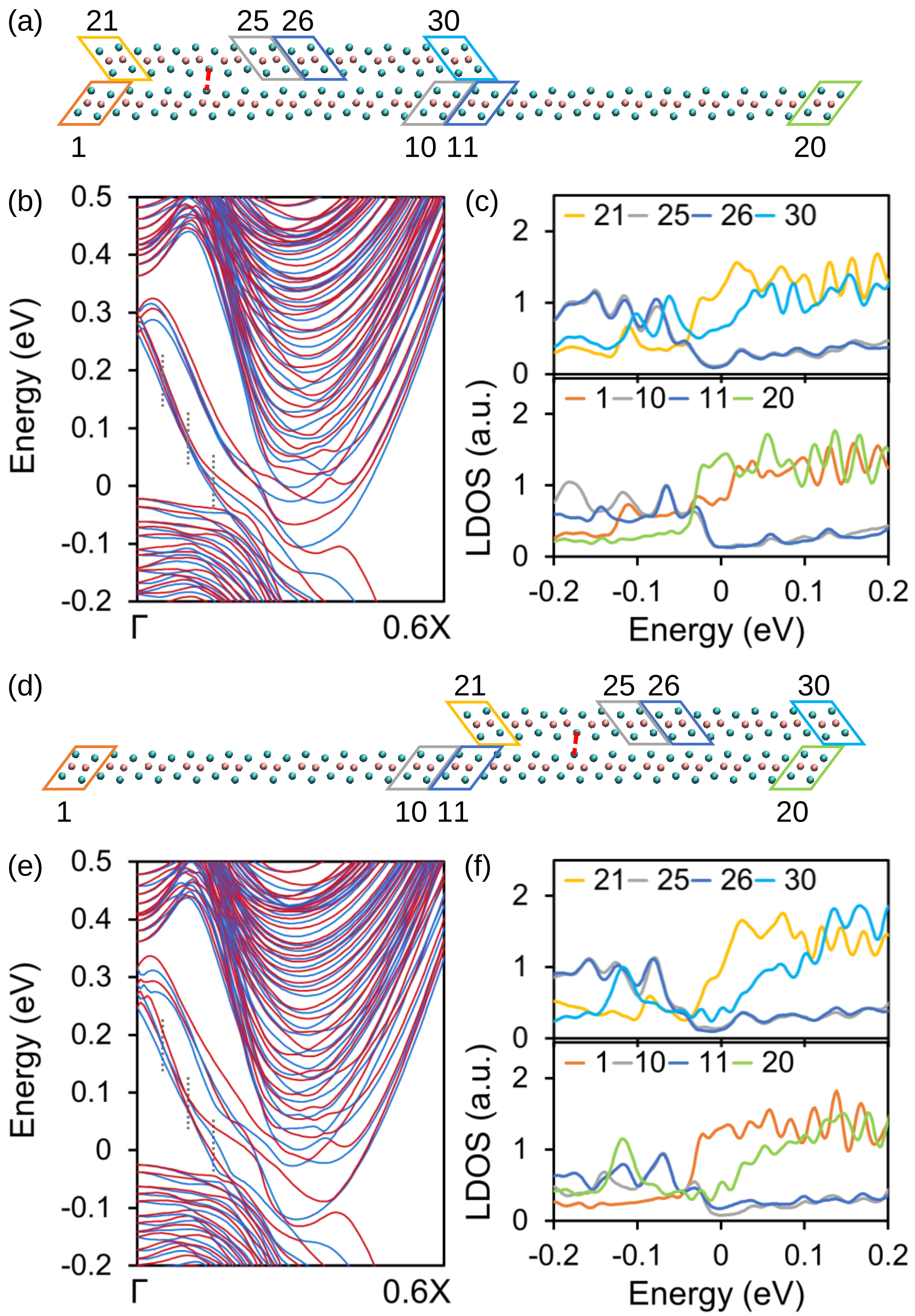

Fig.3.

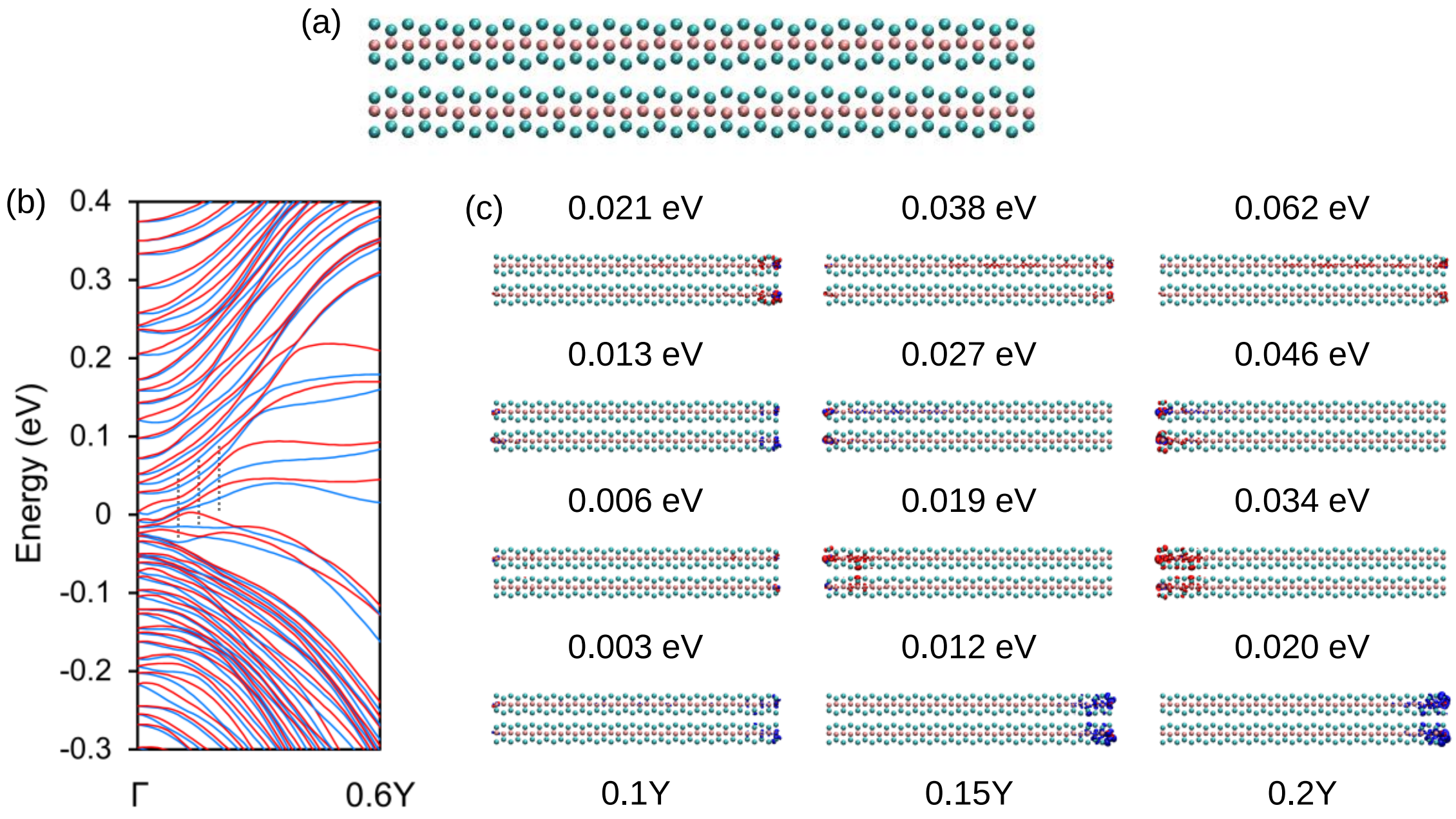

Fig.4.

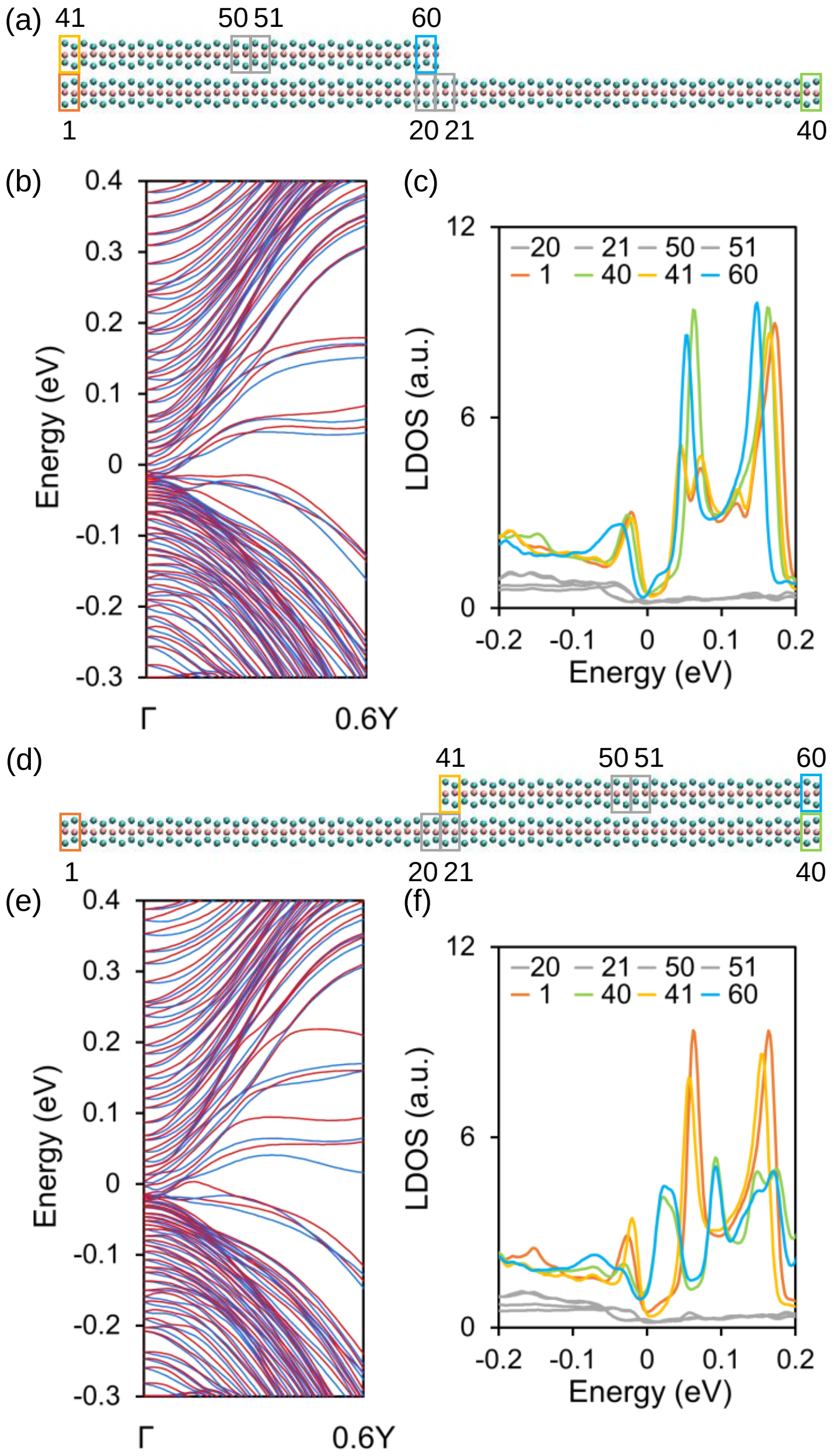

Fig.5.

# Supplementary data: Electronic properties of the steps in bilayer *Td*-$WTe_2$

Mari Ohfuchi[1]*, Akihiko Sekine[1], Manabu Ohtomo[1], Kenichi Kawaguchi[1]

[1]*Fujitsu Research, Fujitsu Limited, Atsugi, Kanagawa 243-0197, Japan*

E-mail: mari.ohfuti@fujitsu.com

### Atomic and Electronic structures of bulk *Td*-$WTe_2$

The atomic structure and the band diagram of bulk *Td*-$WTe_2$ used in this study are presented as follows.

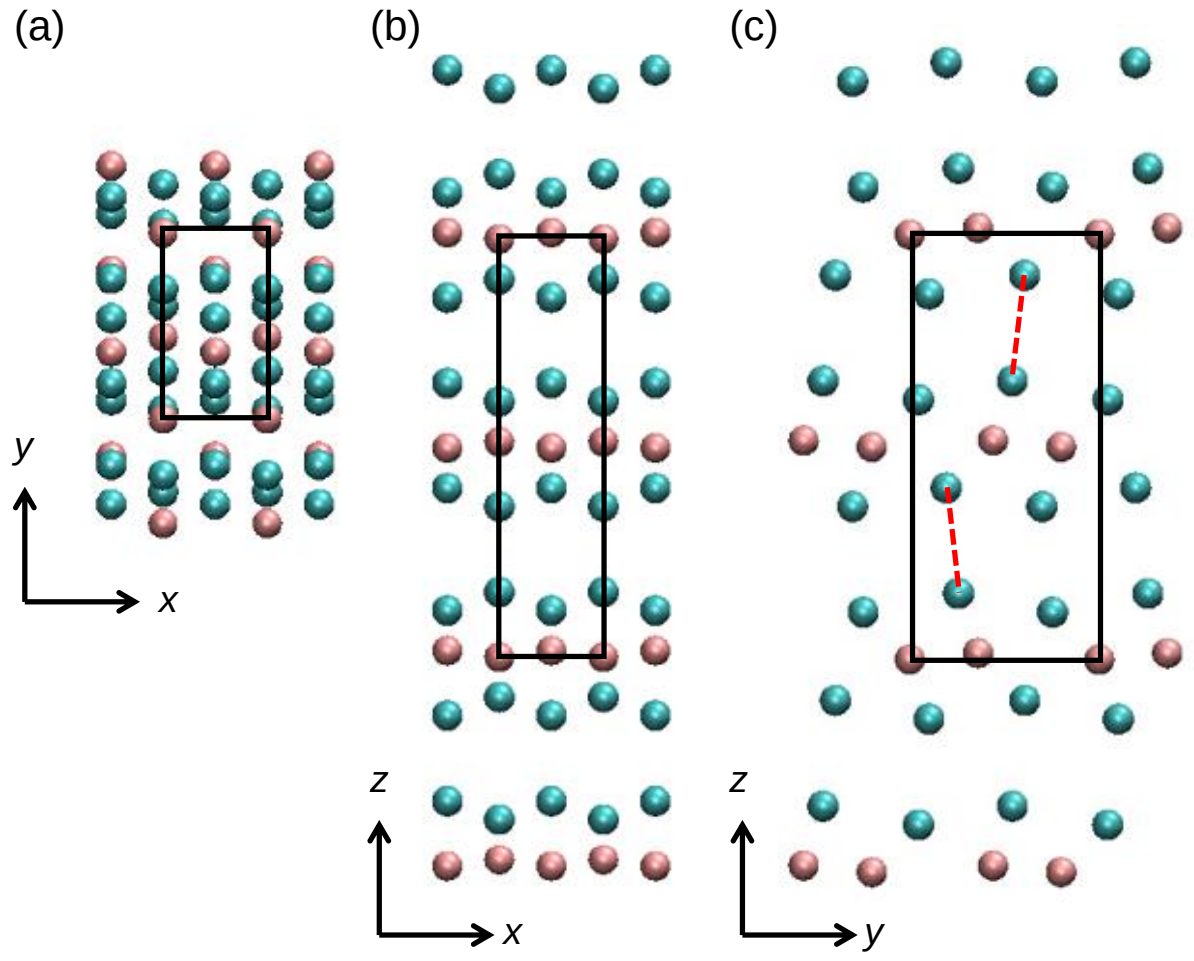

**Fig. S1.** Atomic structure of bulk *Td*-$WTe_2$. The orange and green spheres denote W and Te atoms, respectively, projected toward the (a) *z*, (b) *y*, and (c) *x* directions. The black rectangles indicate the unit cells. The red broken lines are the guidelines representing the alignment of layer stacking.

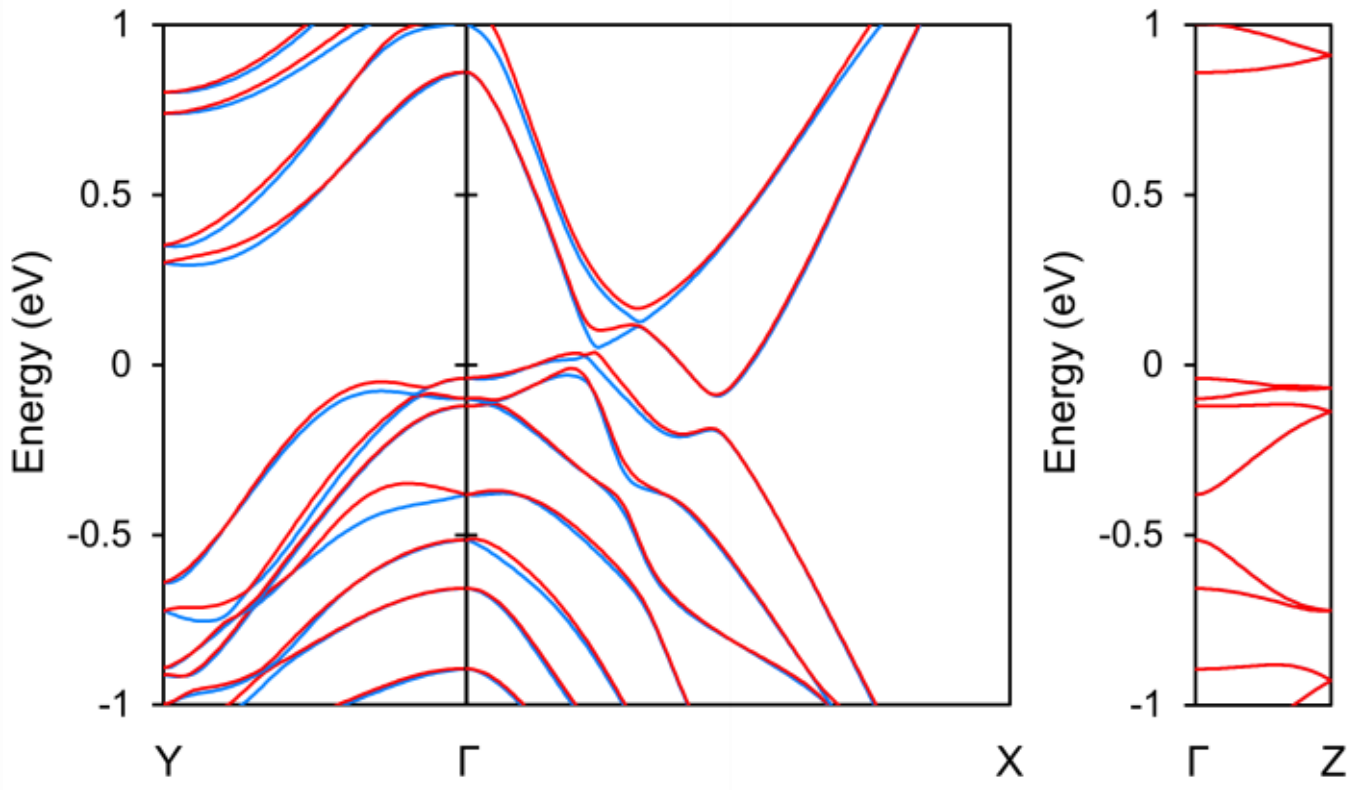

**Fig. S2.** Energy band diagram of bulk *Td*-$WTe_2$. The energy bands are drawn alternately in blue and red in ascending order of the energy to show the spin-orbit coupling splitting.

**Supplementary data for the stripes of monolayer $Td$-$WTe_2$**

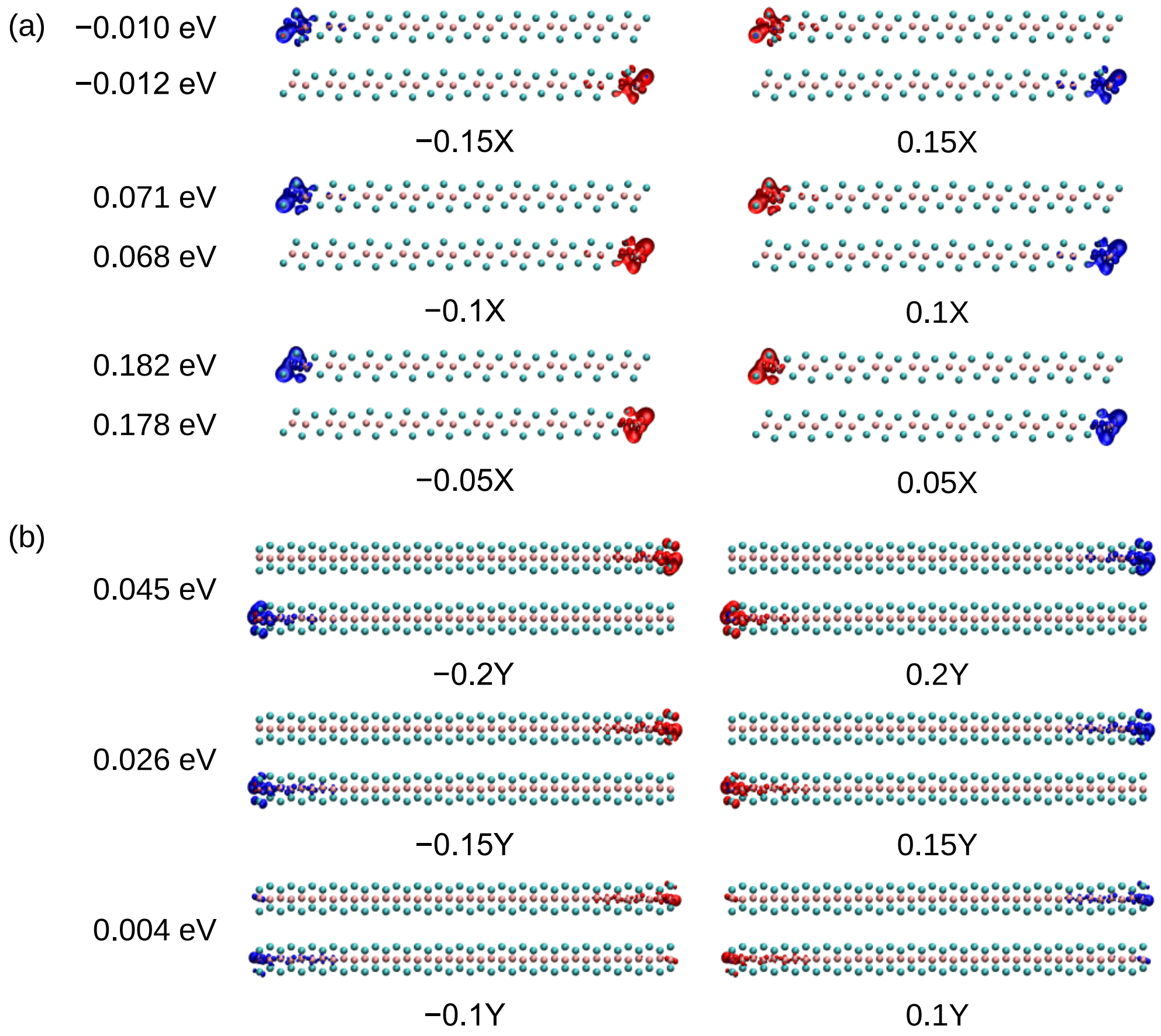

**Fig. S3.** Spin densities for the (a) $x$- and (b) $y$-stripes of monolayer $Td$-$WTe_2$. The spin densities are shown for the points indicated by the broken gray lines in Figs. 1(b) and 1(e). The numerical values show the wavenumber and energy at each point. Two energy levels are degenerated for each point. The blue and red represent up and down spins, respectively. The orange and green spheres denote W and Te atoms, respectively, projected toward the (a) $x$ and (b) $y$ directions.

**Supplementary data for the $x$-stripes of bilayer $Td$-$WTe_2$**

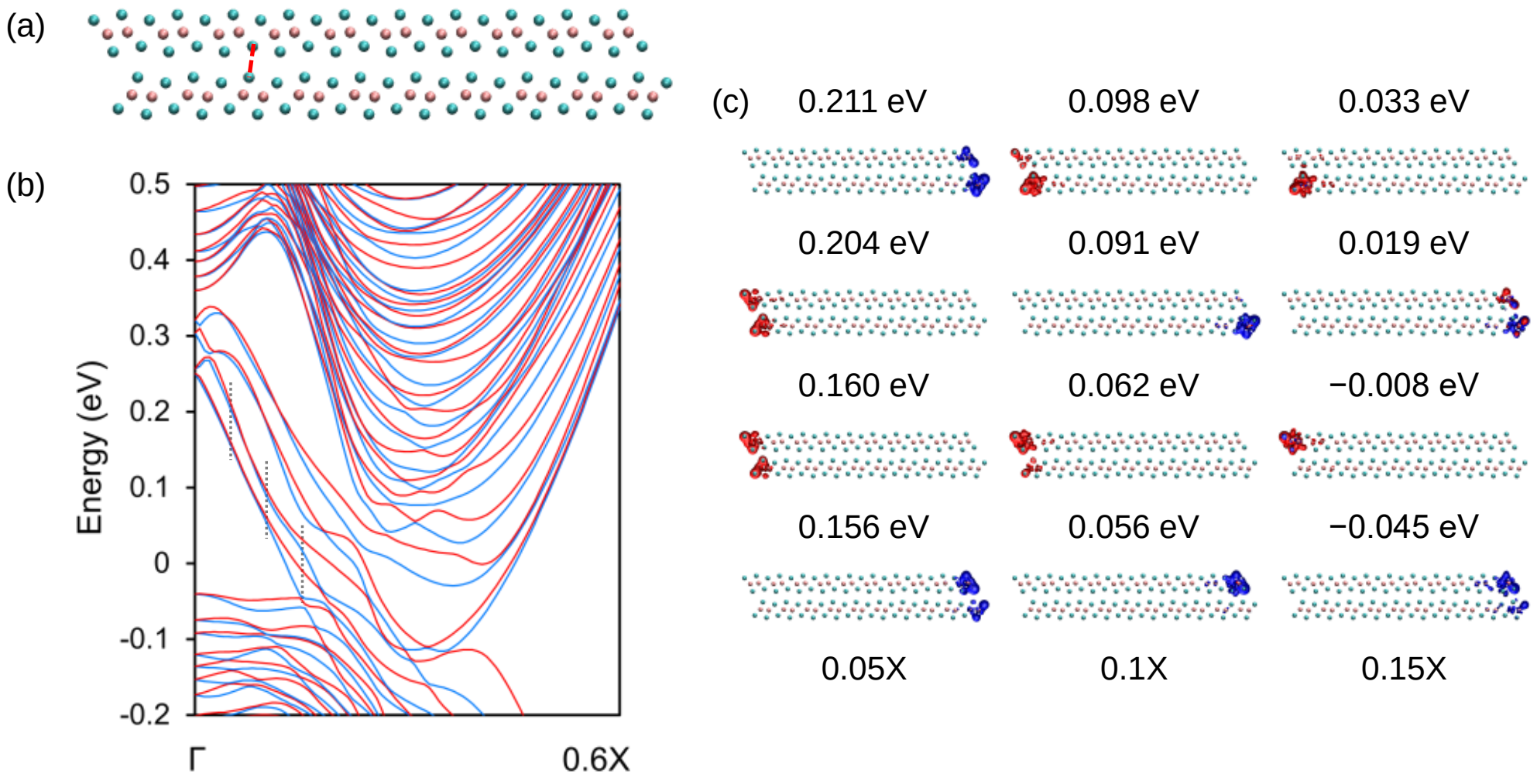

**Fig. S4.** Geometric and electronic properties of the B-$x$-stripe of bilayer $Td$-$WTe_2$: (a) atomic structure, (b) energy

band diagram, and (c) spin densities. The atomic structure is projected toward the $x$ direction. The orange and green spheres denote W and Te atoms, respectively. The broken red line represents the alignment of layer stacking. The energy bands are drawn alternately in blue and red in ascending order of the energy. The spin densities are presented for the points indicated by the broken gray lines in (b). The numerical values show the wavenumber and the energy at each point. The blue and red represent up and down spins, respectively.

## Supplementary data for the steps in bilayer *Td*-$WTe_2$

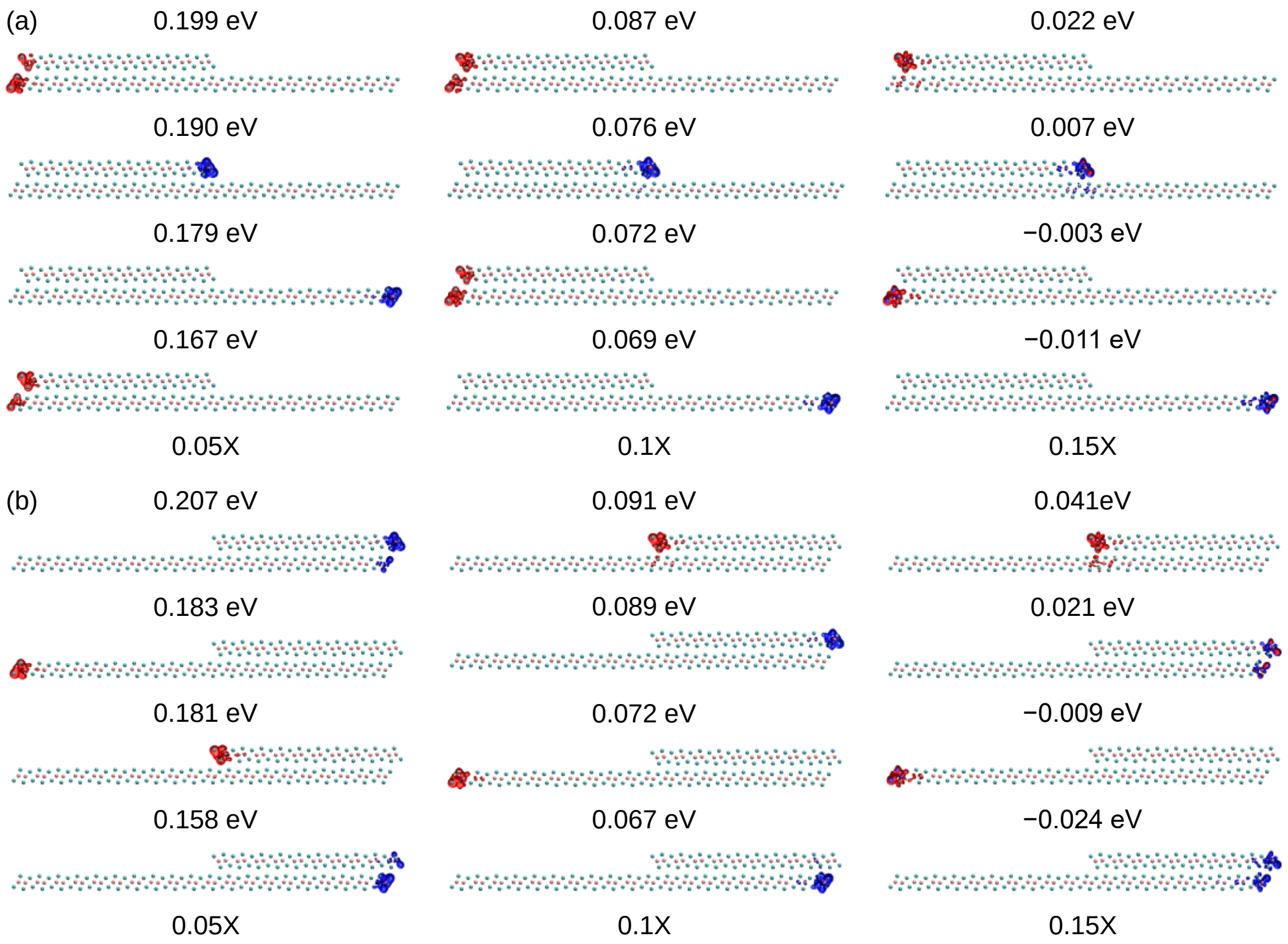

**Fig. S5.** Spin densities for the (a) right and (b) left steps in bilayer *Td*-$WTe_2$. The spin densities at the points indicated by the broken gray lines in Figs. 3(b) and 3(e). The numerical values show the wavenumber and energy at each point. The blue and red represent up and down spins, respectively. The orange and green spheres denote W and Te atoms, respectively. They are projected toward the $x$ direction.

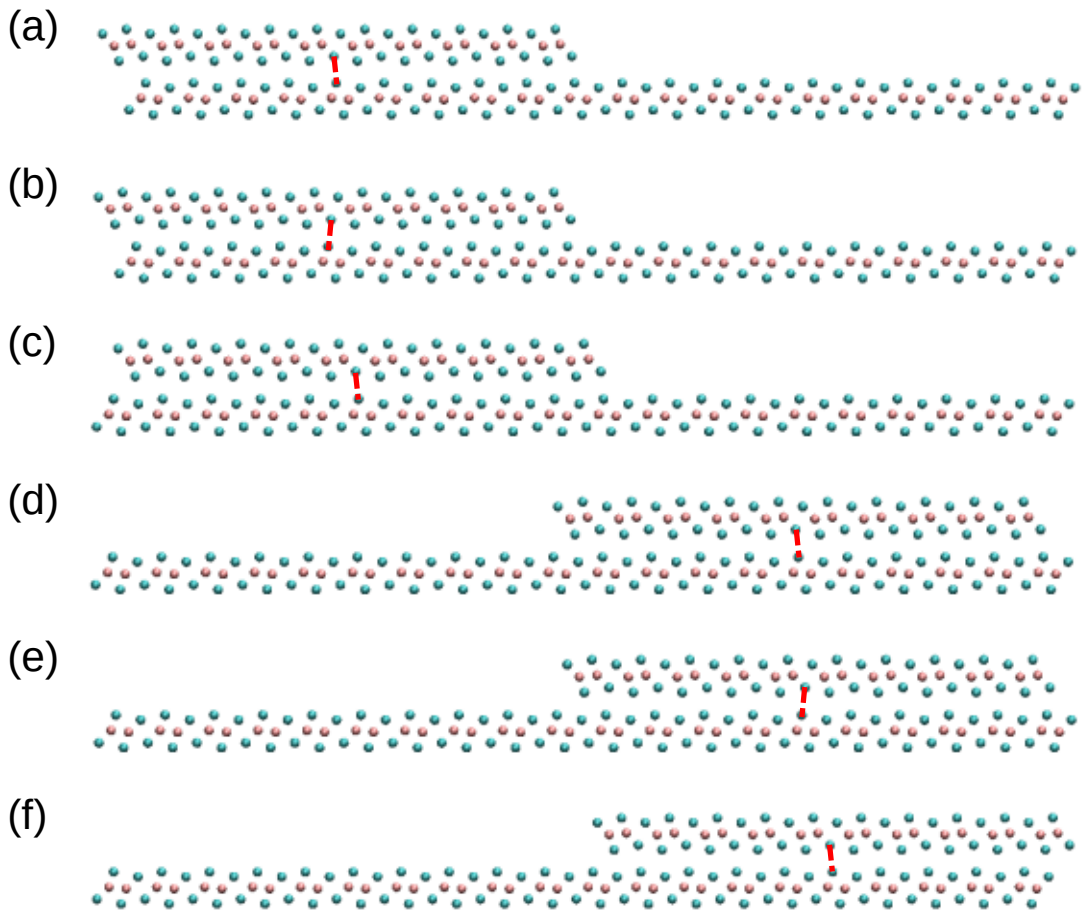

**Fig. S6.** Atomic structures of the *x*-steps in bilayer *Td*-$WTe_2$. The orange and green spheres denote W and Te atoms, respectively. They are projected toward the *x* direction. The red broken lines represent the alignment of layer stacking.

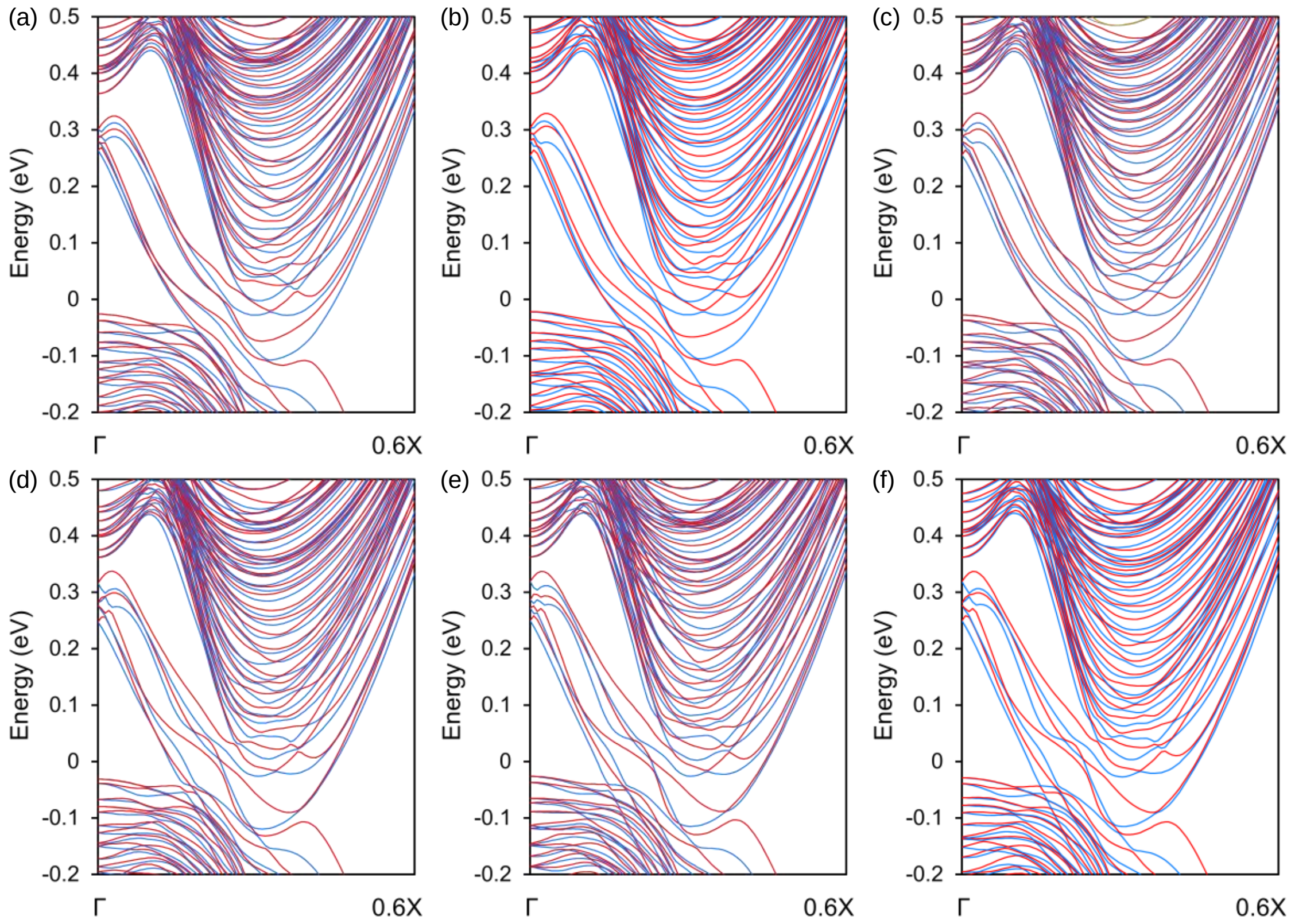

**Fig. S7.** Energy band diagram corresponding to each atomic structure shown in Fig. S6. The energy bands are drawn alternately in blue and red in ascending order of the energy.